\documentstyle[prl,aps,psfig]{revtex}
\begin{document}
\title{Proton--Nucleus Cross Section at High Energies}
\draft
\author{T. Wibig and D. Sobczy\'{n}ska}

\address{Experimental Physics Dept., University of \L odz,
Pomorska 149/153, 90-236 \L odz, Poland}


\maketitle

\begin{abstract}
Cross sections for proton inelastic collision with different nuclei
are described within the Glauber and multiple scattering approximations.
A significant difference between approximate ``Glauber'' formula
and exact calculations with a geometrical scaling assumption 
for very high-energy cross section is shown.
Experimental values of proton--proton cross sections
obtained using extensive air shower data are based on the
relationship of proton--proton and respective proton--air absorption
cross sections. According to obtained results values reported
by the Akeno and Fly's Eye experimental groups are about 10\%
overestimated.
The proper energy dependence of absorption cross section for collisions
with air nuclei is of a great importance for studies of high energy
cosmic rays using the Monte Carlo technique.

\end{abstract}


\section {Introduction}
The rise of the proton-proton cross section (total, inelastic) as the
interaction energy increases is an important feature of the strong
interaction picture. The growth itself is established quite well both
from theoretical and experimental point of view. However the question
how fast do cross sections rise is discussed permanently. A definite answer
is still lacking. Theoretical predictions agree well with one another and
with accelerator data in the region where data exist ($\sqrt{s} \sim 20
\div 2000$ GeV) but they differ above taht level. 
Before the large hadron Collider LHC\cite{lhc} shifts the
direct measurements limit to 10 $\sim$ 14 TeV the only existing information
comes from the cosmic ray extensive air shower (EAS) data.
Two EAS experiments, Akeno \cite{akeno} and Fly's Eye \cite{FE}, gave
estimations of proton--proton total cross section at about $\sqrt{s}
\approx 10^4$ GeV.

The important difference between the collider and EAS proton--proton
cross section measurements is that in fact
the proton--air interactions are involved in the EAS development. Thus
the value which is real measured is the cross section for the interactions
with air nuclei. The value of proton--proton cross section is then obtained
using a theory for nuclei interactions.
In many
recent papers concerning this subject one finds only a brief reference
such as
``calculations have been made in the standard Glauber formalism''
or something very similar \cite{akeno,FE}.

The original Glauber paper \cite{glauber} was published over 40
years ago and it appears that since then some misunderstandings
have arisen. Rather complicated equations for scattering cross sections can
be simplified significantly applying some additional assumptions --- of 
limited
validity. It was pointed out in 1970 
\cite{glaubermat} that some expressions which are most frequently identified
with the nuclear optical model should not be used at least for light nuclei,
yet this still sometimes happens nowadays.

In this paper we compare results of calculations with and without
mentioned simplifications. We will shown that there is quite significant
difference between them. The exact Glauber formalism will be used to evaluate
the proton--proton cross section values from the Akeno and Fly's Eye data.

The paper is organized as follows. In the next section a detail description
of proton--proton scattering, the basis of further nuclei cross section
calculations, is given. In section 3
the proton-nucleus cross section evaluation for 
some commonly used methods is given, and
in section 4 the quantitative results are presented and discussed.

\section {Proton--proton cross sections}
Introducing the impact parameter formalism cross sections can be
described using one, in general complex, function $\chi$
in the form

\begin{eqnarray}
{\sigma}_{\rm tot}~=~2\:\int\:\left[\:1\:-\:{\rm Re}\left (
{\rm e}^{i \chi ({b})}\right)\: \right]
d^2 {\bf b}~~,
\nonumber \\
{\sigma}_{\rm el}~=~\int\:\left| \:1\:-\:{\rm e}^{i \chi ({b})}\:\right| 
^2\:
d^2 {\bf b}~~,
\label{sigsig}
\nonumber \\
{\sigma}_{\rm inel}~=~\int\:1\:-\:\left| \:{\rm e}^{i \chi ({b})}\:\right| 
^2\:
d^2 {\bf b} ~~.
\end{eqnarray}

The phase shift $\chi$ is related to the scattering amplitude
by the two-dimensional Fourier transform

\begin{eqnarray}
1\:-\:{\rm e}^{i\chi({\bf b})}
~=~{{1} \over {2\: \pi \: i}}\int\:{\rm e}
^{-i{\bf b\:t}} S({\bf t}) d^2 {\bf t} ; \nonumber \\
S({t}) ~=~{i \over {2\: \pi \: }}\int\:{\rm e}
^{ i{\bf b\:t}} \left(
1\:-\:{\rm e}^{i\chi({\bf b})} \right)
d^2 {\bf b} .
\label{eq2}
\end{eqnarray}

Using the optical analogy one can interpret the $
1\:-\:{\rm e}^{i\chi({\bf b})}
$ function as a transmission coefficient
for a given impact parameter. Considering two colliding object we can
assume (for pure absorptive potential)

\begin{equation}
\chi({b})~=~i\: \omega(b)~=~i\: K_{ab}\int \: d^2{\bf b'}\:
\rho_a({\bf b})\rho_b({\bf b}\:+\:{\bf b'}) ,
\label{rho}
\end{equation}

\noindent
where $\rho_h$ is a particle's ``opaqueness''
(the matter density integrated
along the collision axis). To some extend the hadronic matter density
could be identified with the charge density measured precisely in the
leptonic-scattering experiments. However, for high-energy hadron--hadron
collisions the real part of the phase shift $\chi$ can be
considerable. Also, the simple interpretation given in Eq.(\ref{rho})
ought to be modified --- at least, by introducing some dependence on the
interaction energy ($s$). Thus the
general phase shift $\chi$ is a two-variable complex function. Two
possible ways of simplifying the situation have been proposed in the 
literature:
the factorization hypothesis (FH) and geometrical scaling (GS). They
can be expressed as

\begin{eqnarray}
\chi(s,\:b) ~=~ i\: \omega(b)\:f(s) ~~~~~~{\rm (FH)} \nonumber \\
\chi(s,\:b) ~=~ i\: \omega \left(b\: / b_0(s)\right) ~~~~~~{\rm (GS)} .
\label{gsfh}
\end{eqnarray}

From the optical point of view, the FH means that the hadron is
getting blacker as the energy increases while the GS means that it is getting
bigger.

The bulk of information about 
the hadron phase-shift function $\chi$ comes from
elastic scattering experiments: more precisely, from the measured differential
elastic cross section. The value of the total cross section can be obtained
from the imaginary part of the forward scattering amplitude using the
optical theorem. Analysis of the elastic data above $\sqrt{s} \sim 20$ GeV
shows that neither assumption given in Eq.(\ref{gsfh}) is realized
exactly (see e.g. Refs.\cite{chouyang,amaldi}).
However, a combination of the two can reproduce the data quite
well.

In this paper the form of $\chi$ is assumed after \cite{men1}
in the form

\begin{equation}
 \chi(s,b)  ~=~(\lambda(s)+i)\: \omega(b,s)
\label{lambdach}
\end{equation}

\noindent
[with $\omega$ defined by Eq.(\ref{rho})].
This follows the original GS idea \cite{buddd} and
differs from the known Martin formula where the ratio of
real to imaginary part of the scattering amplitude depends on the
momentum transfer. A full discussion and some recent
references can be found in \cite{men1}.
In any case, due to a lack of both theoretical and experimental information
about the phase shift, Eq.(\ref{lambdach})
can be treated as a first approximation.
In this paper the parametrization of $\lambda$ used is

\begin{equation}
\lambda(s)~=~{{0.077\: \ln (s/s_0)} \over {
1~+~0.18\: \ln (s/s_0)~+~.015\: \ln ^2 (s/s_0)}}~~~,~~s_0~=~500~{\rm GeV}^2~.
\label{lamb}
\end{equation}

For the $\omega$ energy dependence in Eq.(\ref{gsfh})
the GS is assumed, thus

\begin{equation}
\omega(b,s) ~=~
\omega(\widetilde b ) ~~~~~{\rm with} ~~~~
\widetilde b
~=~  b \:  \left[{{\sigma_{\rm inel}(s_0)}
\over {\sigma_{\rm inel}(s)}} \right]
^{ \frac 1 2 } ,
\label{gsomega}
\end{equation}

\noindent
where $s_0$ is the
center of mass energy for which the detail shape of $\omega$ has been
originally determined.
The accurate data description can be found in Ref.\cite{ws1}
with the ``hadronic matter'' distribution of the form

\begin{equation}
\rho_h({\bf b})~=~\int d z {{ m_h} \over {8 \pi}} {\rm e}^{-m_h {\bf r}}
\label{pdens}
\end{equation}

\noindent
with the coefficients $m_h$ [and $K_{ab}$ in Eq.(\ref{rho})]
adjusted to the hadron--proton elastic scattering data
at $\sqrt{s} \sim 20$ GeV.
The energy dependence of the phase shift is thus introduced by using
the elastic to total cross sections ratio change [$\lambda (s)$ function]
and by the scaling of the interaction impact parameter
which reproduces the increase of cross section as the interaction
energy increases [$\sigma_{\rm inel} (s)$].

The parametrization of ${\sigma_{\rm inel}(s)} $ has been made in the
$\ln ^2 (s)$ form

\begin{equation}
{\sigma_{\rm inel}(s)} = 32.4~-~1.2 \ln (s)~+~0.21 \ln ^2 (s)
\label{inel}
\end{equation}

\noindent
which gives the cross section in mb when $s$ is given in GeV$^2$.
The deviation from the Pomeron-type power-law fit
(see e.g. \cite{blca})
is negligible in the region where the fit can be compared with existing data.
The advantage of using the $\ln ^2$ form is that we do not need to 
concern oruselves about
the violation of the Froissart unitarity bound when using our formula at
very high energies.

Using Eqs.(\ref{pdens}, \ref{lamb}, and \ref{inel}) we
can calculate hadron--nucleon cross sections at any energy of interest
from the point of view of EAS physics.

\begin{figure}
\centerline{\psfig{file=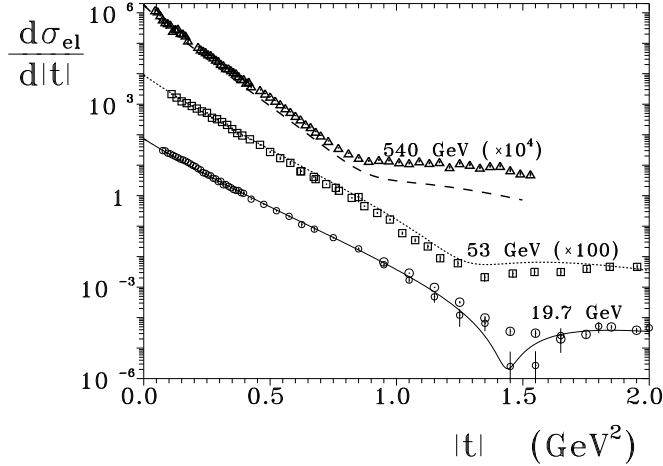,width=10cm}}
\caption{Differential $p$--$p$ elastic cross sections obtained
using the proposed parametrization of $\chi$ for different energies
compared with experimental data from FNAL, ISR and SPS [12].}
\label{elast}
\end{figure}

\begin{figure}
\centerline{\psfig{file=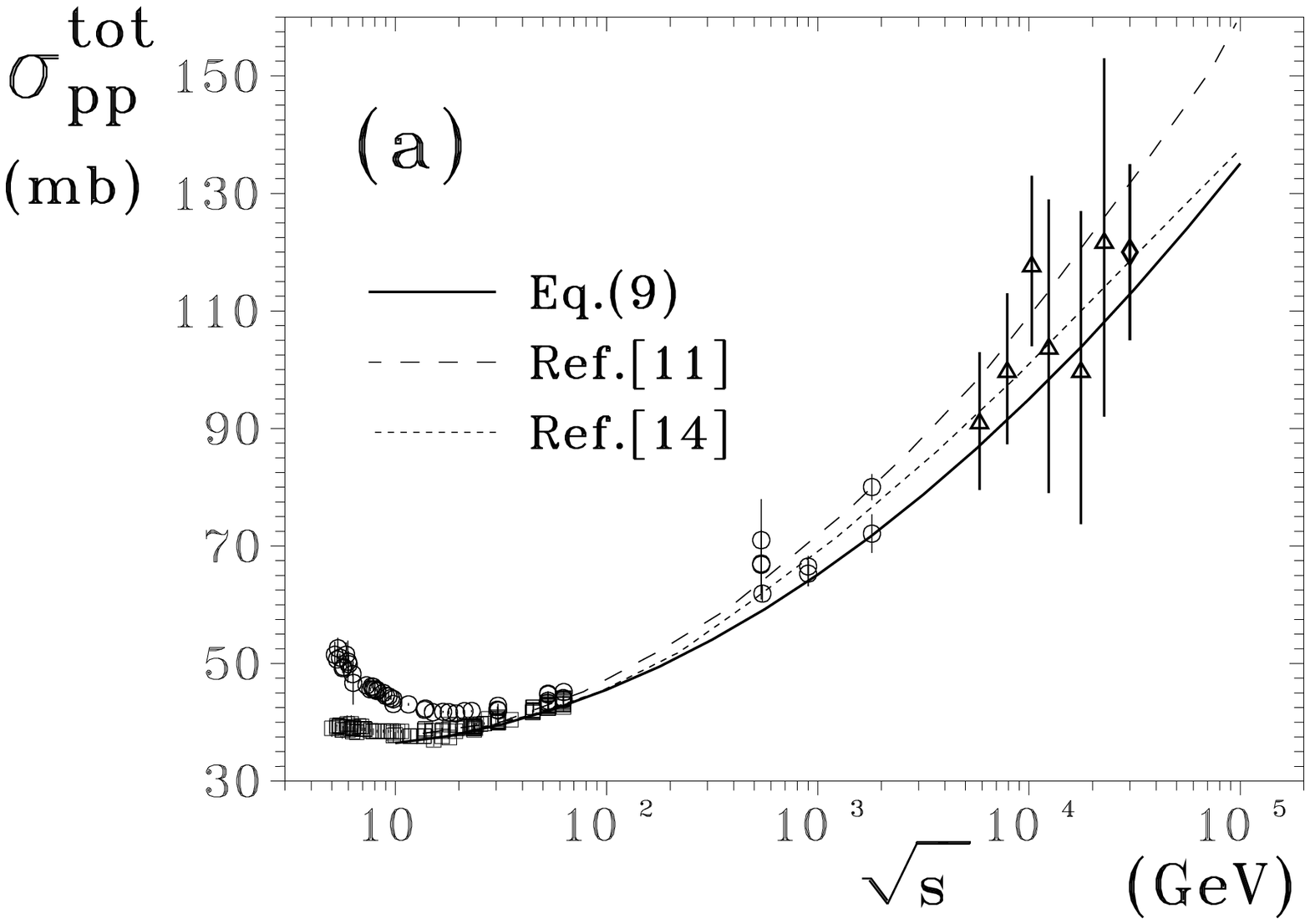,width=9cm}
\psfig{file=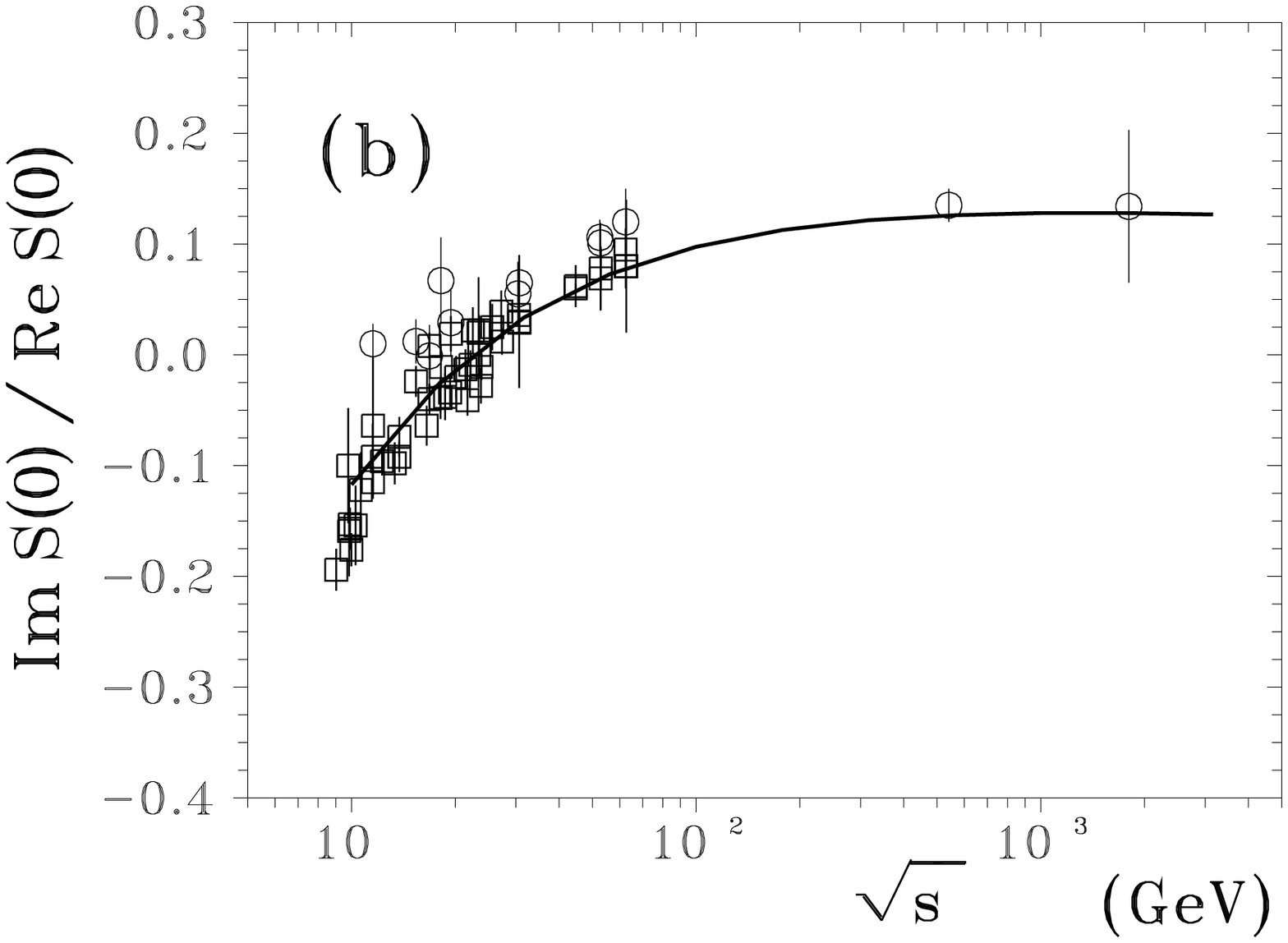,width=9cm}}
\caption{The energy dependence of $p$--$p$
total cross sections (a) and imaginary to real part of elastic forward
amplitude (b)
calculated using the proposed parametrization of $\lambda$
compared with experimental data from [13].
The broken curve shows Block and Cahn fit [11]
and the dotted corve Durand and Pi [14] calculation result.}
\label{sigmas}
\end{figure}

The quality of proposed parametrizations is presented in Figs.\ref{elast}
and \ref{sigmas}.

\section {Nucleon--nucleus cross section calculations}

There are two different approaches
to nuclei cross section calculations. Both give 
similar results, but foundations are quite different. Broadly, it can
be said that the Glauber method is based on the superposition of 
scattering potentials, while 
the other method (which we shall call hereafter the multiple 
scattering method) works on
probabilities.
The multiple scattering approximation is widely presented in \cite{czyzma},
mainly for the elastic scattering of different particles on nuclei, but the
probabilistic formalism can be used to obtain cross sections of other specific
processes.

Of course we do not want to judge which method describes reality better.
The comparison with experimental data is possible in principle,
however, as will be shown, expected differences between these two
approaches are comparable with other effects which has to be known
with appropriate precision (e.g. quasi-elastic scattering).

The description given below is performed for the proton--nucleus interactions,
but the extension to the nucleus--nucleus case is straightforward.

\subsection {Glauber approximation}

For the scattering of particle on the close many-particle system (nucleus),
if each interaction is treated as a two-particle one, the
overall phase shift for incoming wave is a sum of all the two-particle
phase shifts.

\begin{equation}
\chi_A(b,\: \{{\bf d}\})~=~\sum_{j=1}^A \: \chi _j ({\bf b}\: -\: {\bf d}_j)
\label{chiskla}
\end{equation}

\noindent
where $\{{\bf d}\}$ is a set of nucleon positions in the nucleus
(${\bf d}_j$ is a position of the $j$th nucleon in the plane
perpendicular to the interaction axis).
Eq. (\ref{chiskla}) is the essence of the Ref.\cite{glauber}
and in fact defines the Glauber approximation (at least in this paper).

The scattering amplitude is thus given by

\begin{equation}
S(t)~=~{i \over {2 \pi }} \int {\rm e}^{i{\bf t b}} d^2{\bf b}
\int | \psi(\{{\bf d}\} ) | ^ 2 \: \left\{ \: 1 \: - \: {\rm e}
^{i \chi_A(b,\: \{{\bf d}\})}  \right\}
\prod _{j=1}^A d^2 {\bf d}_j~~,
\label{st1}
\end{equation}

\noindent
where $\psi$ describes the wavefunction of the nucleus with nucleons
distributed according to $\{{\bf d}\}$.
If one neglect position correlations of the nucleons and denotes by
$\varrho _j$ each single nucleon density we have

\begin{equation}
|\psi(\{{\bf d}\})|^2~=~\prod_{j=1}^A \varrho_j ({\bf d}_j)
~~~{\rm with}~~~
\int \varrho_j ({\bf r}_j) d^3 {\bf r}~=~1 ~~.
\end{equation}

\noindent
If all interactions can be described by the same phase-shift
function $\chi$ then

\begin{eqnarray}
S(t)={{i} \over {2 \pi }} \int {\rm e}^{i{\bf t b}} d^2{\bf b}
\int
\prod _{j=1}^A \varrho_j ({\bf d}_j)
 \left\{ 1  -  {\rm e} ^{i \sum_{j=1}^{A}\chi({\bf b} - {\bf d}_j)} \right\}
d^2 {\bf d}_j ~=\nonumber \\
=~{{i} \over {2 \pi }} \int {\rm e}^{i{\bf t b}} d^2{\bf b}
\left\{ 1  -   \int  \prod _{j=1}^A \varrho_j ({\bf d}_j)
{\rm e}^{i \chi({\bf b} - {\bf d}_j)} d^2 {\bf d}_j \right\} ~.
\label{st2}
\end{eqnarray}

On the other hand, the scattering process
can be treated as the single collision process
with its own nuclear phase shift $\chi_{\rm opt}(b)$

\begin{equation}
S(t)~=~{{i} \over {2 \pi }} \int {\rm e}^{i{\bf t b}}
\left\{ 1 \: - \: {\rm e}^{i \chi_{\rm opt}(b)} \right\} d^2 {\bf b}~~.
\label{chiop}
\end{equation}

The comparison with Eq.(\ref{st1}) gives

\begin{eqnarray}
 {\rm e}^{i \chi_{\rm opt}(b)}~=~
\int | \psi(\{{\bf d}\})|^2\:
{\rm e}^{i \sum_{j=1}^A \: \chi _j ({\bf b}\: -\: {\bf d}_j)}
\prod _{j=1}^A d^2 {\bf d}_j
~=~\left\langle {\rm e}^{i \chi(b,\: \{{\bf d}\})} \right\rangle~~,
\label{chiopt}
\end{eqnarray}

\noindent
where the $\langle \ \rangle$ means the averaging over all possible
configurations of nucleons $\{ {\bf d} \} $.
To go further with the calculations of $\chi_{\rm opt}$ a commonly used
assumption has to be made. If we assume that the number of scattering centers
($A$) is large and the transparency of the nucleus as a whole remains constant
then

\begin{equation}
\chi_{\rm opt}(b)
~=~i\: \int d^2 {\bf d} \rho_A({\bf d})\:
\left[ 1 - {\rm e}^{i \chi({\bf b} - {\bf d})} \right]~.
\label{exact}
\end{equation}

\noindent
where $\rho_A$ is the
distribution of scattering center (nucleon) positions
in the nucleus ($\sum \varrho_j$).

When the individual nucleon opacity $| 1-{\rm e}^{i \chi(b)} |$ is a very
sharply peaked compared with $\rho_A$ then with the help of the
optical theorem the simple formula can be found

\begin{equation}
\chi_{\rm opt}(b)~=~{1 \over 2}\: \sigma_{pp}^{\rm tot} \left[
\left( {{{\rm Re} S(0)} \over{ {\rm Im} S(0)}} \right) \: + \: i \right]
\rho_A(b)~~.
\label{chig}
\end{equation}

\noindent
Substituting (\ref{chig}) into (\ref{sigsig})
the proton nucleus inelastic cross section is

\begin{equation}
\sigma_{pA}^{\rm inel}~=~
\int d^2 {\bf b}
\left[ 1 - {\rm e}^{- \sigma_{pp}^{\rm tot} \rho_A(b)} \right]
~=~
\int d^2 {\bf b}
\left\{ 1 -
\left[ 1-
\sigma_{pp}^{\rm tot} {\rho_A \over A} \right] ^A \right\}
\label{ginel}
\end{equation}

\noindent
where the last equality holds in the large $A$ limit [
Eq.(\ref{ginel}) cannot be used for $A=1$ to compare results for
$\sigma_{pp}^{\rm inel}$]
This result is often but not quite correctly called
``the Glauber approximation''. As has been shown,
the original Glauber assumption given in Eq.(\ref{chiskla})
has to be supported by small nucleon sizes and a large value of A.

\subsection {Multiple scattering approach}

The $\sigma_{\rm inel}$ given in Eq.(\ref{sigsig}) can be interpreted in the
probabilistic way by identifying the
$[1-|{\rm e}^{i\chi({\bf b})}|^2]$
term as the
probability of inelastic scattering at impact parameter ${\bf b}$. 
This can be
extended to the interaction with nucleus in a straightforward way.
If we denote this probability by $P({\bf b})$
%
%
and nucleons in a nucleus $A$ are distributed according to $\rho_A$
then the averaged probability of inelastic interaction with one of 
the nucleons is

\begin{equation}
\overline{P}_A({\bf b})~=~
\int d^2 {\bf d} \ {{\rho_A({\bf b})} \over A} \ P({\bf b} - {\bf d}).
\end{equation}

The inelastic cross section with the whole nucleus is then

\begin{equation}
\sigma_{pA}^{\rm inel}~=~
\int d^2 {\bf b} \left\{
1-
\left[ 1-
\overline{P_A({\bf b})}\right]
^A
\right\}~~.
\end{equation}

In the multiple scattering picture the point-nucleon approximation can be
also introduced simplifying the cross section formula. If one puts
$P({\bf b})~=~\delta^2({\bf b})\:\sigma_{pp}^{\rm inel}$ then

\begin{equation}
\overline{P}_A({\bf b})~=~
{{\rho_A({\bf b})} \over A} \sigma_{pp}^{\rm inel},
\end{equation}

\noindent
what leads to

\begin{equation}
\sigma_{pA}^{\rm inel}~=~
\int d^2 {\bf b} \left\{ 1 - \left[ 1-
\sigma_{pp}^{\rm inel} {{\rho_A ({\bf b})} \over A}\right]^A \right\}
\label{minel}
\end{equation}

The above equation has a very similar form to 
Eq.(\ref{ginel}) but the difference is also quite clear.

It is interesting to note that Eq.(\ref{minel})
is often called the ``Glauber approximation'' too (see e.g.
\cite{gs2}).

\subsection{Comparison with the exact Glauber formula}

Both approaches discussed above lead to slightly different formulae
for an inelastic proton--nucleus cross section. To see how big the difference
is we have calculated respective cross sections using exactly the same
procedures, nucleus shapes and proton--proton cross sections.
It is interesting to compare results for different nucleus masses and
incoming proton energies.

The energy dependence in the discussed approximation is introduced only
via the proton-proton cross sections change. Respective formulae
have been given in section 2

The distributions of the nucleon
position in the nuclei of the form given in the FRITIOF
interaction model \cite{frit} have been used

\begin{equation}
\varrho_j~\sim~\left\{ \matrix{ {\textstyle
\left\{\: {1+\exp
\left[ \left( r- r_0 A^{1/3} \right)/C \right] \:
 } \right\} }
^{-1}
& \ \ \ \  {\rm for} & A > 16 \cr
{\textstyle
\left[ 1+ {{A-4} \over 6} \left( r \over d \right) ^2 \right]
\exp \left( - {r \over d}^2 \right)} & \ & A \leq 16
}
\right.
\end{equation}

\noindent
with the parameters given there, except for $A=4$ where we have used
the so-called ``parabolic Fermi distribution'' \cite{gs2} and for
lighter ($A=2,3$) nuclei, where the simple Gaussian was used. The minimum
allowable distance between two nucleons was introduced (0.8 fm) which
modify the ``initial'' (uncorrelated) distribution rather strongly,
especially for very light nuclei, so the detail shape of the light
nuclei uncorrelated density distribution is not necessary for our
purposes.

\begin{figure}
\centerline{\psfig{file=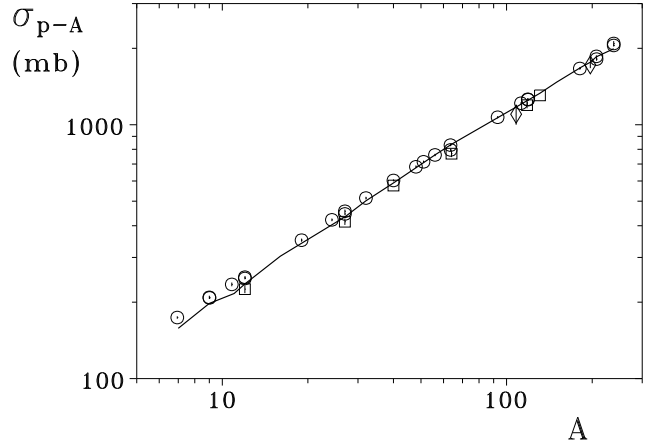,width=10cm}}
\caption{
Inelastic cross sections measured at the laboratory proton
momentum of $\sim$ 200 GeV/$c$ compared with our calculation results
as a function of target atomic mass. Data from Ref.[18].}
\label{paxsect}
\end{figure}

The validity of the nucleus description used is shown in Fig.\ref{paxsect}
where the
proton--nucleus inelastic cross section data measured at low energies
are compared with our calculations. The interaction energy is about
the same that it was used as the reference energy $s_0$ for the
proton ``hadronic matter density distribution'' ($\omega$) estimation,
so the energy dependence of all proton--proton cross section parameters
are not involved here.

\begin{figure}
\centerline{
\psfig{file=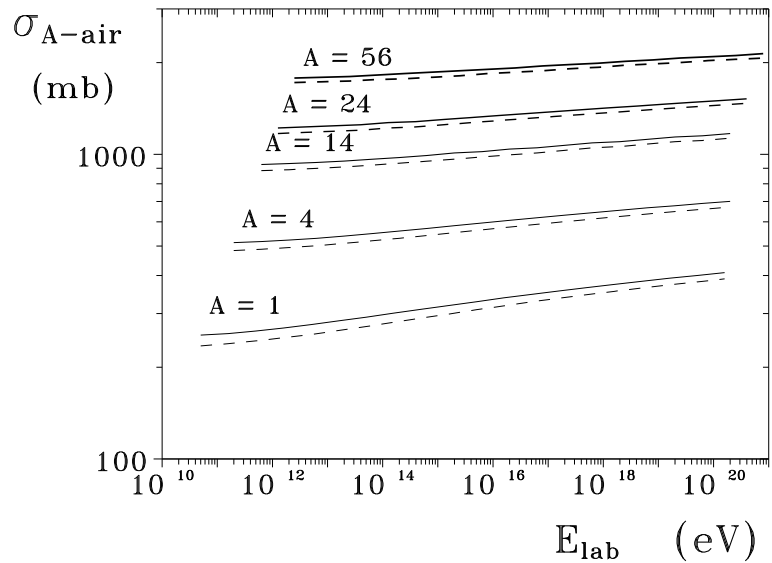,width=12cm}}
\caption{Cross section of collisions
of different nuclei with the ``air nucleus'' calculated
with simplified Glauber (full lines)
and multiple scattering (broken lines) approximations for different
interaction energies.}
\label{xsecc1}
\end{figure}

The comparison between the simplified Glauber [Eq.(\ref{ginel})] and
multiple scattering [Eq.(\ref{minel})] results is given in
Fig.~\ref{xsecc1} for proton--nucleus as well as for nucleus--nucleus
interactions. We have chosen here the ``air nucleus'' as a target and
five different projectiles which represent main components of primary
cosmic ray mass spectrum. The difference is not very significant. The energy
dependence is similar and the systematic shift is rather constant as 
maight be expected. It should be remembered that for light nuclei (especially
for protons) Eq.(\ref{ginel}) looses its physical basis.

\begin{figure}
\centerline{
\psfig{file=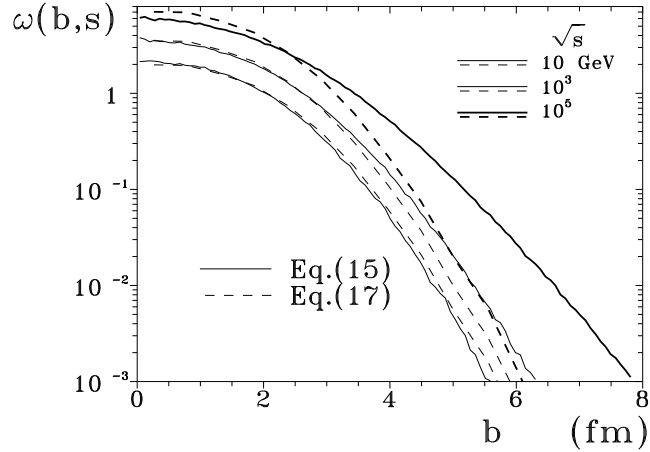,width=10cm}}
\caption{Nitrogen nucleus profile functions ($\chi_{p-{\rm N}}$)
obtained using the
exact Glauber formula with geometrical scaling (full curves)
and factorization hypothesis
(simplified Glauber and multiple scattering) (broken curves) for
different interaction energy (per proton--nucleon collision).}
\label{ddis}
\end{figure}

The important point of this paper is to show how the point-nucleon
approximation changes the results.
Both formulae given in Eqs.(\ref{minel}) and (\ref{ginel}) are in fact in
agreement with the factorization hypothesis for individual nucleon--nucleon
$\chi$ function [Eq.(\ref{gsfh})], according to which nucleons get
blacker
as the interaction energy increases. Our analysis, presented in section 2,
strongly favoured the geometrical scaling which treats nucleons
as getting bigger. Nucleus profiles obtained using exact Glauber formula
[Eq.(\ref{exact})] differ from the
$A \sigma_{pp}^{\rm inel} {\rho_A \over A}$ 
suggested by
Eq.(\ref{minel}) [$A \sigma_{pp}^{\rm tot} {\rho_A \over A}$ in
Eq.(\ref{ginel})]. The difference can be seen in Fig.\ref{ddis}.

\begin{figure}
\centerline{\psfig{file=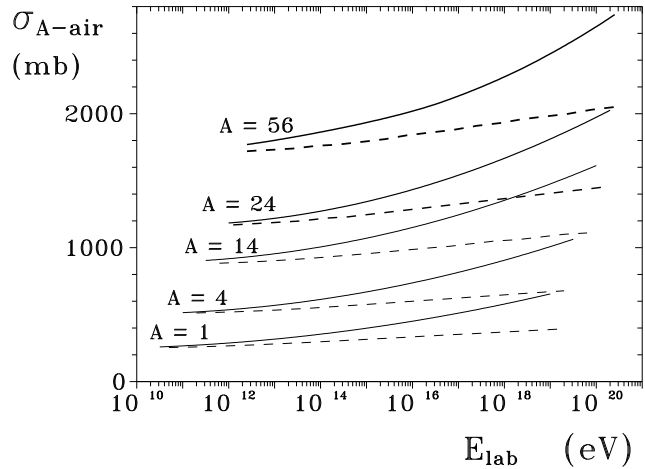,width=10cm}}
\caption{
Cross section of collisions
of different nuclei with the ``air nucleus'' calculated
using the exact Glauber formula with geometrical scaling
(full curves) and simplified formulas (broken curves)
as a function of interaction energy.}
\label{finxsec}
\end{figure}

A significant change of the nucleus size have to influence the value
of inelastic cross section. Fig.\ref{finxsec} shows the
change in inelastic cross section of proton--nucleus with the
interaction energy calculated using geometrically scaled and
factorized nucleus profiles.

As can be seen, the difference at very high energies is remarkable.

\section {Proton--proton cross section from cosmic ray data.}

Results presented in the previous section indicate the importance of
re-examination of the proton--proton cross section estimation based
on proton--air data measured in EAS experiment.

\begin{figure}
\centerline{\psfig{file=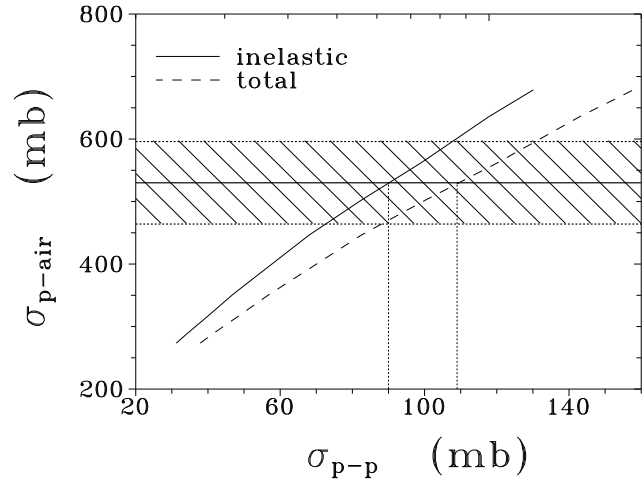,width=10cm}}
\caption{The relationship between inelastic proton--air cross section
and the value of proton--proton cross section (inelastic - full curve
and total - broken curve)
calculated using exact Glauber formula with geometrical scaling.
Full horizontal line represents the value measured in Fly's Eye [3]
experiment (dashed area shows 1$\sigma$ bounds).}
\label{pppair}
\end{figure}

The conversion from proton--air to proton--proton cross section
presented in Fig.\ref{pppair} is obtained using the exact Glauber
formalism.

The original Fly's Eye estimation of proton--proton total cross section
given in \cite{FE} is 120 mb, yet according to results given in
Fig.\ref{pppair} it is 109 mb. The same procedure 
was applied to the Akeno data and all their proton--proton cross sections
also appear to be about 10\% overestimated.

High-energy points plotted in Fig.\ref{sigmas}(a) which are often 
reproduced in the
literature are taken from the original works.
They agree quite well with the two phenomenological descriptions plotted
there. The approximation of the cross section rise used in this paper
[Eq.(\ref{inel})] agrees with the Block and Cahn and with the Durand and Pi
parametrizations in the low-energy region, but falls
below the original EAS proton--proton data points.

\begin{figure}
\centerline{\psfig{file=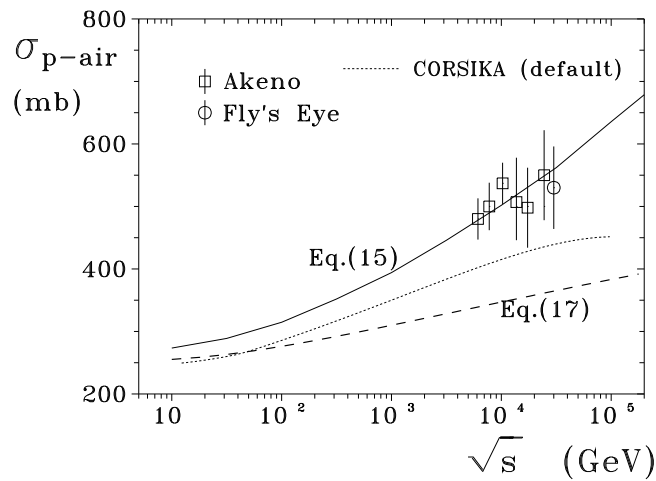,width=10cm}}
\caption{Inelastic proton--air cross sections
calculated using exact Glauber formula with geometrical scaling
as a function of interaction energy (per nucleus--nucleus interaction).
Experimental points are from Akeno and Fly's Eye experiments
(squares and the circle, respectively). Results of
calculations with the ``simplified Glauber'' formula with the
same proton--proton cross section energy dependence are given by the
broken curve. The default CORSIKA proton--air
cross section is shown for a comparison by the dotted curve.}
\label{pair}
\end{figure}

In Fig.\ref{pair} the calculated proton--air cross section energy
dependence is given. The full curve represents result 
obtained using the exact Glauber formula [Eq.(\ref{chiopt})] and 
proton--proton
phase shift $\chi$ function described in section 2 [with the proton--proton
cross section rise defined by Eq.(\ref{inel})and shown in Fig.\ref{sigmas}
by the full curve]. 
The outcome of the
``simplified Glauber'' approach [Eq.(\ref{chig})] is given for a comparison
by the broken curve.

As can be seen, the proposed cross section rise gives very good agreement
with EAS measurements at about $\sqrt{s} \sim 10^4$ GeV.

Comparison of our results with outcomes from some of the
most popular shower development codes (see e.g. \cite{corsika}
with different models used there) shows that
energy dependence of nucleus--air cross sections used there are
rather flat, which perhaps suggests that a ``simplified Glauber'' have
been used in cross section calculations (see the dotted curve
in Fig.\ref{finxsec}). The relatively correct, fast,
rise of the SIBYLL \cite{sib} model proton--air cross section
is obtained with ``simplified Glauber'' formalism, but with a quite
extraordinary rise of the total proton--proton cross section at high energies
(reaching about 150 mb at $\sqrt{s} \sim 3 \cdot 10^{4}$ GeV).

\section{Summary}
We have shown that
the geometrical scaling hypothesis with the exact Glauber formalism
gives the value of proton--proton total cross section
at about 30 TeV slightly (10\%) smaller than that reported in original
Fly's Eye and Akeno papers. The fit for $\sigma_{pp}^{\rm inel}$
given in Eq.(\ref{inel}) leads to $\sigma_{pp}^{\rm tot}$ shown in
Fig.\ref{sigmas}(a) by the full curve and correctly reproduces
$\sigma_{p-air}$ cross section measured by EAS experiments as it is shown in 
Fig.\ref{pair}.

The rise of the $\sigma_{A-air}$ predicted by
the geometrical scaling hypothesis with the exact Glauber formalism
is significantly faster than that which can be obtained using
simplified formulas (Fig.\ref{finxsec}).
This can change the physical conclusions based on
Monte Carlo simulations of the EAS development at very high energies.

\end{document}